\documentclass[twocolumn,secnumarabic,amssymb, nobibnotes, aps, prd, superscriptaddress]{revtex4-1}
\usepackage{amsmath}
\usepackage{graphicx}
\usepackage{mathtools}
\usepackage{epstopdf}

\newcommand{\env}[1]{\texttt{#1}}
\newcommand{\angstrom}{\mbox{\normalfont\AA}}
\begin{document}

\title{Nonlinear Response of Bilayer Graphene at Terahertz Frequencies}%

\author{Riley McGouran}%
%\email[REVTeX Support: ]{rileymcgrn@gmail.com}
\affiliation{Department of Physics, Engineering Physics and Astronomy, Queen’s University, Kingston, Ontario K7L 3N6, Canada}
\author{Ibraheem Al-Naib}%
%\email[REVTeX Support: ]{rileymcgrn@gmail.com}
\affiliation{ Biomedical Engineering Department, College of Engineering, University of Dammam, Dammam 31441, Kingdom of Saudi Arabia.
}
\author{Marc M. Dignam}%
%\email[REVTeX Support: ]{rileymcgrn@gmail.com}
\affiliation{Department of Physics, Engineering Physics and Astronomy, Queen’s University, Kingston, Ontario K7L 3N6, Canada}
\date{August 16, 2016}%

\begin{abstract}
A density-matrix formalism within the length gauge is developed for the purpose of calculating the nonlinear response of intrinsic bilayer graphene at terahertz frequencies. Employing a tight-binding model, we find that interplay between the interband and intraband dynamics leads to strong harmonic generation at moderate field amplitudes. Specifically, we find that at low temperature (10 K), the reflected field of undoped suspended bilayer graphene exhibits a third harmonic amplitude that is 30\% of the fundamental in the reflected field for an incident 1 THz single-cycle pulse with a field amplitude of 1.5  kV/cm. More interestingly, we find that as the central frequency of the incident radiation is increased, the third harmonic amplitude also increases; reaching a maximum of 53\% for an incident frequency of 2 THz and amplitude of 2.5 kV/cm.
\end{abstract}

\maketitle
%\tableofcontents

\section{Introduction}
Commencing in 2004 with the first observation of free standing graphene - along with the measurement of its electronic properties \cite{Geim:graphene2004}- there has been an abundance of both experimental and theoretical work performed on graphene and its multi-layered counterparts \cite{Novo:graphene2005, geim:riseofgraphene2007, nilsson2006electronic, mccann2013electronic, bowlan2014ultrafast, paul2013high}. The impetus behind this research arises from the intriguing electrical, mechanical, and optical properties possessed by graphene. Many of the intriguing properties of monolayer graphene (MLG) are also present in bilayer graphene (BLG). These include a zero-gap energy dispersion, a very high carrier mobility,  high thermal conductivity, and the ability to tune electric properties by adjusting the carrier density through gating \cite{Novo:graphene2005, Morozov:GiantIntrinsic2008}. 

Despite their similarities, BLG differs from MLG in a number of ways. One important distinction is found in the density of states of both systems. Close to the \textit{Dirac} points - the points within the first Brillouin zone at which the conduction and valence bands touch - one can show that although the interband matrix elements for both systems are essentially the same, in BLG, the number of states within 10 meV of the Dirac point is approximately 30 times larger than in MLG. This difference in the density of states will have an effect on the interband transitions of the two systems. In order to probe this energy range one needs to probe BLG with radiation that has frequencies in the terahertz range. 

As with MLG, the absorption of optical and THz radiation by BLG can be understood in terms of interband and intraband transitions. The energies of photons at THz frequencies are ideally suited to the study of intrinsic (undoped) BLG, where the Fermi level is at the Dirac point. At these frequencies, interband transitions can take place near the Dirac point, while the injected electrons and holes are subsequently strongly driven within their respective bands. As we shall see, the ability of THz radiation to induce both intraband and interband dynamics is fundamental in the emergence of a strong nonlinear response in BLG. 
%The THz response is dependent on the Fermi level, scattering mechanisms, %sample temperature, incident field amplitude, and central frequency of the %pulse. 

Earlier studies on MLG have suggested the presence of a strong nonlinearity at both optical and THz  frequencies \cite{mikhailov2007non, mikhailovnonlin2009, hendry2010coherent, ishikawa2010nonlinear, wright2009strong}. Theoretical studies of BLG have indicated a strong nonlinear effect in the THz to far-infrared regime, whereby a moderate electric field can result in third harmonic generation at room temperature \cite{ang2010nonlinear}. Experiments have been performed with the intent of observing the nonlinear THz response of graphene, specifically third harmonic generation. Although it has been observed by using a 45-layer sample, it has not yet been successfully observed in MLG or BLG \cite{bowlan2014ultrafast, paul2013high}.

Almost all experimental investigations of the nonlinear response of graphene have observed a response that is dominated by intraband dynamics, due to the high doping level of the samples. The large Fermi energy of the doped system diminishes the interband current due to Pauli blocking. For more moderate Fermi energies - tens of meV - the interband dynamics can make a contribution that is dependent on the field amplitude. In recent theoretical work on MLG, it was shown that if the Fermi level is reduced to within a few meV of the Dirac point, the magnitude of the interband current becomes comparable to the intraband current, resulting in the presence of a strong nonlinearity \cite{al2014high}. Specifically, it was found that odd harmonics of the THz field should be generated. As this interplay between interband and intraband dynamics is a result of the zero-gap band structure, we also expect to see a similar interplay in the carrier dynamics in BLG.

In this paper, we employ a four-band tight-binding method to model the intraband and interband dynamics of undoped suspended bilayer graphene in response to a single-cycle THz pulse in the range of 1 to 5 THz. We use this model to calculate the dependency of the third harmonic response of BLG on the graphene temperature and the central frequency of the pulse. The current densities and the generated harmonics are calculated numerically for two values of the temperature, 10 and 100 K. We find that the ratio of the amplitude of the third harmonic to the fundamental in the reflected field is reduced by an order of magnitude as the temperature is increased from 10 to 100 K. Finally, we examine the nonlinear response at a number of central THz frequencies. We find that as the central frequency increases from 1 to 5 THz, there is an increase in the ratio of the third harmonic to the fundamental; reaching a maximum at a central frequency of 2 THz.

The paper is organized as follows. In section \ref{sec:2}, we present the derivation of our theoretical model. A tight-binding method is used to obtain dipole matrix elements within the length gauge. The matrix elements are then used in a four-band density matrix formalism in order to calculate the interband and intraband current densities. In section \ref{sec:3}, we present both the linear and  nonlinear THz response of numerical simulations for undoped BLG at different temperatures and central THz frequencies. The conclusions are presented in section \ref{sec:4}.

\section{Theory}\label{sec:2}
The calculations that we perform are based on a theoretical approach employing a density-matrix formalism in the length gauge.  A nearest-neighbor tight-binding model is used to treat the $\pi$-electrons in the graphene, which are taken to provide the conduction electrons only \cite{sarma2011electronic}. 
\subsection{Energy Bands}
\begin{figure}[h]
\centering
\includegraphics[width=0.35\textwidth]{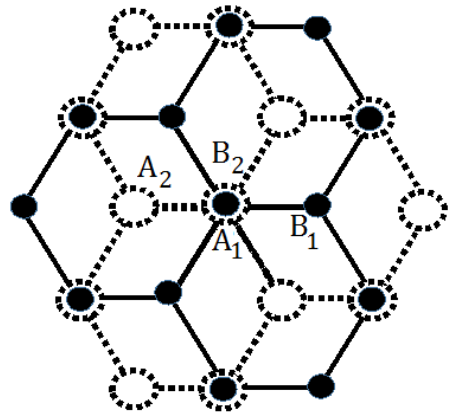}
\caption{Schematic diagram of the crystal structure of BLG and the labelling conventions for atoms in the unit cell. A sites are in white, while B sites are in black\label{fig:lattice2}.}
\end{figure}

The tight-binding expression for the Bloch states is given by 
\begin{equation}\begin{aligned}\psi_{n\mathbf{k}}(\mathbf{r}) & =A_{n}\left(\mathbf{k}\right)\sum_{i}\sum_{\mathbb{\mathbf{R}}}C_{i}^{n}\left(\mathbf{k}\right)\varphi_{pz}\left(\mathbf{r}-\mathbf{R-r}_{i}\right)e^{i\mathbf{k}\cdot\mathbf{R}},\end{aligned}
\end{equation}
where $A_{n}\left(\mathbf{k}\right)$ is a normalization factor, $n$ labels the conduction and valence bands, and the sum is over the Bravais lattice vectors $\mathbf{R}$. The sublattice coefficients, $C_{i}^{n}\left(\mathbf{k}\right)$,
 are associated with the four carbon atoms within the unit cell; the $\varphi_{pz}\left(\mathbf{r}\right)$ are the $2p_{z}$ orbitals of carbon. The index $i$ indicates a sum over the basis vectors $\mathbf{r}_{A_{1}},\,\mathbf{r}_{B_{1}},\,\mathbf{r}_{A_{2}},\,\mathbf{r}_{B_{2}}$, which give the position of sublattice sites $A_{1}$ and $B_{1}$  in the top layer, and $A_{2}$ and $B_{2}$ in the bottom layer (see Fig. \ref{fig:lattice2}). Explicitly, they are given by $\mathbf{r}_{A_{1}}=0$, $\mathbf{r}_{B_{1}}=a_{o}\widehat{\mathbf{x}}$, $\mathbf{r}_{A_{2}}=-a_{o}\widehat{\mathbf{x}}$ and $\mathbf{r}_{B_{2}}=0$.

 The vector of the sublattice coefficients is given by $$\Bigl|\mathbf{k}_{n}\Bigr\rangle^{\dagger}=\left(C_{A_{1}}^{n*}\left(\mathbf{k}\right),\, C_{B_{1}}^{n*}\left(\mathbf{k}\right),\, C_{A_{2}}^{n*}\left(\mathbf{k}\right),\, C_{B_{2}}^{n*}\left(\mathbf{k}\right)\right).$$ Within the basis of these vectors, the secular equation for BLG can be expressed as
\begin{equation}
\textrm{det}\left[\begin{array}{cccc}
-\epsilon_{n\mathbf{k}} & f\left(\mathbf{k}\right)t_{\parallel} & 0 & t_{\perp}\\
f\left(\mathbf{k}\right)^{*}t_{\parallel} & -\epsilon_{n\mathbf{k}} & 0 & 0\\
0 & 0 & -\epsilon_{n\mathbf{k}} & f\left(\mathbf{k}\right)t_{\parallel}\\
t_{\perp} & 0 & f\left(\mathbf{k}\right)^{*}t_{\parallel} & -\epsilon_{n\mathbf{k}}
\end{array}\right]=0,\label{eq:sec}
\end{equation}
\vspace{3mm}
with the eigenvalue of band $n$ given by $\epsilon_{n\mathbf{k}}$. 

The \textit{intralayer} hopping energy, $t_{\parallel}$, and the \textit{interlayer} hopping energy, $t_{\perp}$, are approximately equal to 3.03 eV and 0.3 eV, respectively \cite{malard2007probing}. The function $f\left(\mathbf{k}\right)\equiv(1+e^{-i\mathbf{k}\cdot\mathbf{a}_{1}}+e^{-i\mathbf{k}\cdot\mathbf{a}_{2}})$, is a result of the nearest neighbor intralayer electron hopping, where the $\mathbf{a}_i$ are the primitive translation vectors of graphene, given explicitly by \begin{equation}\begin{aligned}\mathbf{a}_{1} & =\frac{3a_{o}}{2}\widehat{\mathbf{x}}+\frac{\sqrt{3}a_{o}}{2}\widehat{\mathbf{y}},\end{aligned}
 \begin{aligned}\mathbf{a}_{2} & =\frac{3a_{o}}{2}\widehat{\mathbf{x}}-\frac{\sqrt{3}a_{o}}{2}\widehat{\mathbf{y}}.\end{aligned}\end{equation} Here $a_{o}$ is the nearest-neighbour separation $(a_{o}\simeq1.42\, \angstrom)$, not the length of the primitive vectors.
 
Equation (\ref{eq:sec}) can be solved exactly to yield expressions for the eigenvalues of BLG. In order of increasing energy, these are given by
\begin{equation}\begin{aligned}E_{v_{1}}=-\frac{\tilde{\epsilon}\left(\mathbf{k}\right)+t_{\perp}}{2} &  & E_{v_{2}}=-\frac{\tilde{\epsilon}\left(\mathbf{k}\right)-t_{\perp}}{2}\\
\\
E_{c_{1}}=\frac{\tilde{\epsilon}\left(\mathbf{k}\right)-t_{\perp}}{2} &  & E_{c_{2}}=\frac{\tilde{\epsilon}\left(\mathbf{k}\right)+t_{\perp}}{2},
\end{aligned}\label{eq:eig}\end{equation} where the first two expressions represent the valence bands, the second two the conduction bands and where we define $\tilde{\epsilon}\left(\mathbf{k}\right)=\sqrt{t_{\perp}^{2}+4|f\left(\mathbf{k}\right)|^{2}.}$

\begin{figure}[h]
\centering
\includegraphics[width=0.5\textwidth]{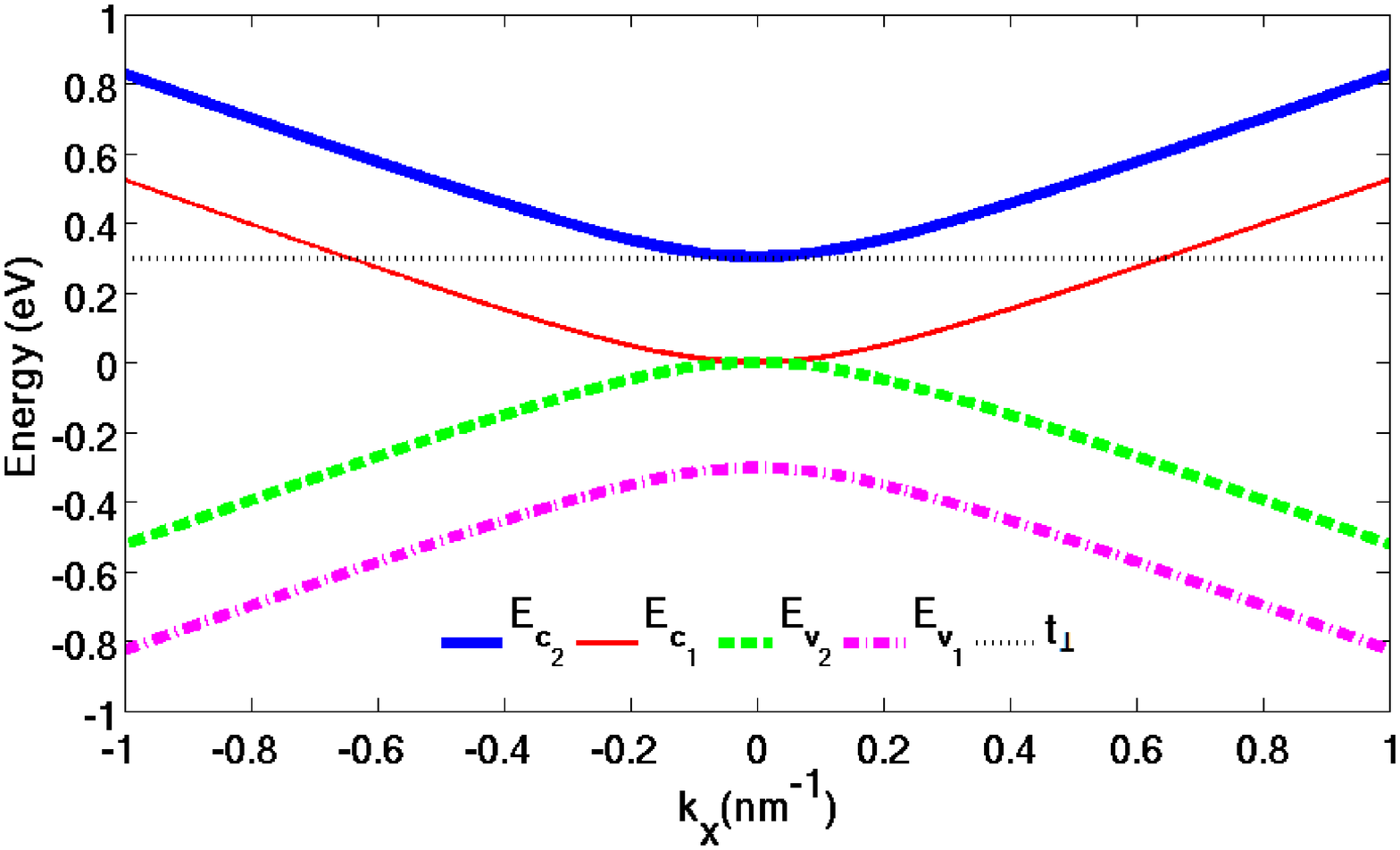}
\caption{Band structure of unbiased bilayer graphene. Lower energy bands touch at the Dirac point; higher energy bands are separated by $2t_{\perp}$\label{fig:1}.}
\end{figure}

Due to the symmetry between the sublattices, the conduction and valence states in graphene are degenerate at two \textit{Dirac points}, given by: \begin{equation}\begin{aligned}\mathbf{K}a_{o} & =\frac{4\pi}{3\sqrt{3}}\widehat{\mathbf{y}},\\
\mathbf{K}^{\prime}a_{o} & =\frac{8\pi}{3\sqrt{3}}\widehat{\mathbf{y}}.
\end{aligned}\end{equation}
For energies within a few hundred meV of the Dirac points, we can expand the crystal momentum around the Dirac points as $\mathbf{k}=\mathbf{K}+\delta\mathbf{k}$ and $\mathbf{k}^{\prime}=\mathbf{K}^{\prime}+\delta\mathbf{k}$, where $\delta\mathbf{k}=k_{x}\hat{\mathbf{x}}+k_{y}\hat{\mathbf{y}}$. With this expansion, we can express the band energy near both Dirac points as $\tilde{\epsilon}\left(\mathbf{k}\right)$ as \begin{equation}\tilde{\epsilon}\left(\mathbf{K}+\delta\mathbf{k}\right)\approx\sqrt{t_{\perp}^{2}+\alpha k^{2}},\end{equation}where $k$ is the magnitude of the crystal momentum $k\equiv|\delta\mathbf{k}|$, $\alpha=4\hbar^{2}v_{F}^{2}$, where $v_{F}=3a_{o}t_{\perp}/2\hbar$ is the Fermi velocity. The energy dispersion given in Eq. (\ref{eq:eig}) for the four bands is shown in Fig. \ref{fig:1} for $k$-vectors near the $\mathbf{K}$-Dirac point.

Using the expressions for the eigenvalues in Eq. (\ref{eq:eig}), we may also solve for the corresponding eigenstates exactly, to obtain

\begin{equation}
\Bigl|\mathbf{k}_{v_{1}}\Bigr\rangle=\frac{1}{2}\left(\begin{array}{c}
-\left(\frac{\tilde{\epsilon}+t_{\perp}}{\tilde{\epsilon}}\right)^{1/2}\\
\left(\frac{\tilde{\epsilon}-t_{\perp}}{\tilde{\epsilon}}\right)^{1/2}e^{-i\chi}\\
-\left(\frac{\tilde{\epsilon}-t_{\perp}}{\tilde{\epsilon}}\right)^{1/2}e^{i\chi}\\
\left(\frac{\tilde{\epsilon}+t_{\perp}}{\tilde{\epsilon}}\right)^{1/2}
\end{array}\right),
\end{equation}

\begin{equation}
\Bigl|\mathbf{k}_{v_{2}}\Bigr\rangle=\frac{1}{2}\left(\begin{array}{c}
\left(\frac{\tilde{\epsilon}-t_{\perp}}{\tilde{\epsilon}}\right)^{1/2}\\
-\left(\frac{\tilde{\epsilon}+t_{\perp}}{\tilde{\epsilon}}\right)^{1/2}e^{-i\chi}\\
-\left(\frac{\tilde{\epsilon}+t_{\perp}}{\tilde{\epsilon}}\right)^{1/2}e^{i\chi}\\
\left(\frac{\tilde{\epsilon}-t_{\perp}}{\tilde{\epsilon}}\right)^{1/2}
\end{array}\right),
\end{equation}

\begin{equation}
\Bigl|\mathbf{k}_{c_{1}}\Bigr\rangle=\frac{1}{2}\left(\begin{array}{c}
-\left(\frac{\tilde{\epsilon}-t_{\perp}}{\tilde{\epsilon}}\right)^{1/2}\\
-\left(\frac{\tilde{\epsilon}+t_{\perp}}{\tilde{\epsilon}}\right)^{1/2}e^{-i\chi}\\
\left(\frac{\tilde{\epsilon}+t_{\perp}}{\tilde{\epsilon}}\right)^{1/2}e^{i\chi}\\
\left(\frac{\tilde{\epsilon}-t_{\perp}}{\tilde{\epsilon}}\right)^{1/2}
\end{array}\right),
\end{equation}

\begin{equation}
\Bigl|\mathbf{k}_{c_{2}}\Bigr\rangle=\frac{1}{2}\left(\begin{array}{c}
\left(\frac{\tilde{\epsilon}+t_{\perp}}{\tilde{\epsilon}}\right)^{1/2}\\
\left(\frac{\tilde{\epsilon}-t_{\perp}}{\tilde{\epsilon}}\right)^{1/2}e^{-i\chi}\\
\left(\frac{\tilde{\epsilon}-t_{\perp}}{\tilde{\epsilon}}\right)^{1/2}e^{i\chi}\\
\left(\frac{\tilde{\epsilon}+t_{\perp}}{\tilde{\epsilon}}\right)^{1/2}
\end{array}\right),
\end{equation}
where $ e^{i\chi\left(\mathbf{k}\right)}=f\left(\mathbf{k}\right)/\left\vert f\left(\mathbf{k}\right)\right\vert$, and we have supressed the explicit $\mathbf{k}$-dependencies for simplicity. 

As discussed in previous studies \cite{al2014high, virk2007semiconductor, aversa1995nonlinear}, modelling the nonlinear response of semiconductors in a limited band model is only expected to be accurate if one uses the length gauge. If instead the velocity gauge is employed, unphysical low-frequency divergences arise in the nonlinear response. As these divergences affect the response in the THz regime, one must develop sum rules in order to remove them. Moreover, trying to obtain the response to higher and higher order in the field, these sum rules become analytically intractable. Consequently, when using a limited basis of bands, the length gauge is advantageous. Hence, to model the THz interaction of BLG, we employ the length gauge Hamiltonian, expressed as $H=H_{0}-e\mathbf{r}\cdot\mathbf{E}$, where $H_{0}$ is the full Hamiltonian of
unperturbed BLG, $e = -|e|$ is the charge of an electron, $\mathbf{r}$ is the electron position vector, and $\mathbf{E}(t)$ is the THz electric field at the graphene. Given that we only consider normally incident plane waves, the field is uniform over the graphene sheets.

The carrier dynamics in BLG are calculated by solving the equations of motion for the density matrix in the basis of the Bloch states describing the four bands of BLG. For these, we require the matrix elements of the Hamiltonian between the various Bloch states: 

\begin{equation}\Bigl\langle\mathbf{k}_{n}\Bigr|H\Bigl|\mathbf{k}_{m}^{\prime}\Bigr\rangle=E_{n}(\mathbf{k})\delta(\mathbf{k}-\mathbf{k}^{\prime})\mathbf{\delta}_{nm}-e\Bigl\langle\mathbf{k}_{n}\Bigr|\mathbf{r}\Bigl|\mathbf{k}_{m}^{\prime}\Bigr\rangle\cdot\mathbf{E}(t).\label{eq:9}
 \end{equation}

The matrix elements of $\mathbf{r}$ between Bloch states can be shown to be given by \cite{al2014high, aversa1995nonlinear}

\begin{equation}\left\langle n,\mathbf{k}\right.\left|\mathbf{r}\right.\left|m,\mathbf{k}^{\prime}\right\rangle =\delta(\mathbf{k}-\mathbf{k}^{\prime})\mathbf{\xi}_{nm}(\mathbf{k})+i\delta_{nm}\mathbf{\nabla_{k}}\delta(\mathbf{k}-\mathbf{k}^{\prime}),
 \label{eq:1}\end{equation} where the connection elements are defined as

\begin{equation}\mathbf{\xi}_{nm}(\mathbf{k})=\frac{\left(2\pi\right)^{2}i}{\Omega}\int_{\Omega}d^{3}\mathbf{r}u_{n,\mathbf{k}}^{\ast}\left(\mathbf{r}\right)\mathbf{\nabla}_{\mathbf{k}}u_{m,\mathbf{k}}\left(\mathbf{r}\right),
\end{equation} where $\Omega$ is the volume of a unit cell and  $u_{n\mathbf{k}}(\mathbf{r})$ is the periodic part of the Bloch function. These connection elements have been calculated using our tight-binding wave function. Ignoring the overlap of atomic wave functions on different atoms, near the $\mathbf{K}$-Dirac point the interband connection elements can be shown to be given by

\begin{equation}
\begin{aligned}\xi_{c_{1}v_{2}}\left(\mathbf{K}+\delta\mathbf{k}\right) & =\left(\frac{\tilde{\epsilon}+t_{\perp}}{2\tilde{\epsilon}}\right)\frac{\widehat{\mathbf{\theta}}}{k}\\
\xi_{c_{2}v_{1}}\left(\mathbf{K}+\delta\mathbf{k}\right) & =\left(\frac{\tilde{\epsilon}-t_{\perp}}{2\tilde{\epsilon}}\right)\frac{\widehat{\mathbf{\theta}}}{k}\\
\xi_{c_{1}v_{1}}\left(\mathbf{K}+\delta\mathbf{k}\right) & =-\frac{i\sqrt{\alpha}t_{\perp}}{2\tilde{\epsilon}^{2}}\hat{\mathbf{k}}\\
\xi_{c_{2}c_{1}}\left(\mathbf{K}+\delta\mathbf{k}\right) & =-\frac{\sqrt{\alpha}}{2\tilde{\epsilon}}\widehat{\mathbf{\mathbf{\theta}}}\\
\xi_{c_{2}v_{2}}\left(\mathbf{k}\right) & =\xi_{c_{1}v_{1}}^{*}\left(\mathbf{k}\right)\\
\xi_{v_{2}v_{1}}\left(\mathbf{k}\right) & =\xi_{c_{2}c_{1}}\left(\mathbf{k}\right),
\end{aligned}\label{eq:zeta}\end{equation} where $\delta\hat{\mathbf{k}}=cos(\theta)\hat{\mathbf{x}}+sin(\theta)\hat{\mathbf{y}}$ and $\hat{\theta}=-sin(\theta)\hat{\mathbf{x}}+cos(\theta)\hat{\mathbf{y}}$ are, respectively, the radial and angular unit vectors in cylindrical coordinates with the origin at the $\mathbf{K}$-Dirac point. Around the $\mathbf{K}^{\prime}$-Dirac point, we find that all the connection elements change sign except for those with components in the $\hat{\mathbf{k}}$-direction. Explicitly, we have
\begin{equation}
\begin{aligned}
\xi_{c_{1}v_{2}}\left(\mathbf{K}^{\prime}+\delta\mathbf{k}\right) & =-\xi_{c_{1}v_{2}}\left(\mathbf{K}+\delta\mathbf{k}\right)\\
\xi_{c_{2}v_{1}}\left(\mathbf{K}^{\prime}+\delta\mathbf{k}\right) & =-\xi_{c_{2}v_{1}}\left(\mathbf{K}+\delta\mathbf{k}\right)\\
\xi_{c_{2}c_{1}}\left(\mathbf{K}^{\prime}+\delta\mathbf{k}\right) & =-\xi_{c_{2}c_{1}}\left(\mathbf{K}+\delta\mathbf{k}\right)\\
\xi_{c_{1}v_{1}}\left(\mathbf{K}^{\prime}+\delta\mathbf{k}\right) & =\xi_{c_{1}v_{1}}\left(\mathbf{K}+\delta\mathbf{k}\right).
\end{aligned}\label{eq:zeta2}
\end{equation}

 Similarly, we can show that all the intraband connection elements - which are identical to the Berry connections of the respective bands \cite{chang1996berry, xiao2010berry} -  are zero, i.e.
\begin{equation}\xi_{nn}\left(\mathbf{k}\right)\equiv\mathcal{A}_{n}\left(\mathbf{k}\right)=0.\end{equation} This is in agreement with the idea that the Berry phase of BLG does not have to be simply $2\pi$, but rather can be an integer multiple of $2\pi$ \cite{park2011berry}. The vanishing Berry connection leads directly to a zero Berry curvature for each of the four bands; which is expected due to inversion symmetry present in BLG. 

The dynamic equations for the density matrix elements can be found using the approach of Aversa and Sipe \cite{aversa1995nonlinear}, which was used in similar work on MLG \cite{al2014high}. These equations are given by
\begin{equation}
\begin{alignedat}{1}\frac{d\rho_{nm}(\mathbf{k})}{dt} & =i\frac{e}{\hbar}\mathbf{E}(t)\cdot\sum_{l}(\mathbf{\xi}_{nl}(\mathbf{k})\rho_{lm}(\mathbf{k})-\rho_{nl}(\mathbf{k})\mathbf{\xi}_{lm}(\mathbf{k}))\\
 & -\frac{e}{\hbar}\mathbf{E}(t)\cdot\nabla_{\mathbf{k}}\rho_{nm}(\mathbf{k})-i\omega_{nm}(\mathbf{k})\rho_{nm}(\mathbf{k})\\
 & -\frac{\rho_{nm}\left(\mathbf{k}\right)-\delta_{nm}\rho_{nm}^{eq}(\mathbf{k})}{\tau_{nm}},
\end{alignedat}
 \end{equation} where $n,m=\{c_{1},c_{2},v_{1},v_{2}\}$, $ \omega_{nm}(\mathbf{k})=\omega_{n}(\mathbf{k})-\omega_{m}(\mathbf{k})=(E_{n}(\mathbf{k})-E_{m}(\mathbf{k}))/\hbar,$ and $\rho_{nm}^{eq}(\mathbf{k})=f_{n}(\mathbf{k},T)\delta_{nm}$ is the carrier population in equilibrium when $n=m$, and is zero otherwise; where $f_{n}(\mathbf{k},T)$ is the Fermi-Dirac distribution at temperature $T$. In our numerical work, we model the vacancy populations rather than the valence band electrons to allow us to only include states near the Dirac point, which greatly reduces computation time. 

Previous studies have shown that the scattering times in graphene are on the order of tens of femtoseconds \cite{bowlan2014ultrafast, paul2013high}, therefore to accurately model the THz response we must take into account scattering processes. At the low carrier densities considered in this work, carrier-carrier scattering is expected to be relatively unimportant, and the dominant scattering processes will be defect scattering and electron-phonon scattering. To account for these mechanisms in our model, we treat scattering phenomenologically. We introduce an interband decoherence time, $\tau_{nm}$, for the interband coherences, $\rho_{nm}(\mathbf{k})$, where $n\neq m$. We assume the decoherence time to be independent of $\mathbf{k}$. The populations, $\rho_{nn}(\mathbf{k})$, we take to relax back to Fermi-Dirac thermal distributions, $f_{n}(\mathbf{k},T)$, with relaxation times, $\tau_{n}$. Since the interaction with THz radiation induces interband transitions, as the simulation proceeds, we adjust the temperature of the Fermi-Dirac distribution so that the carriers relax to the instantaneous carrier populations. Also, we neglect interband relaxation as it has been found that the intraband scattering times are much shorter than the time taken for conduction band electrons to relax to the valence band \cite{tielrooij2013}.

We employ a direct computational approach to solve the above equations, wherein we put $\mathbf{k}$ on a grid and step through time using a Runge-Kutta algorithm. To facilitate this, we make use of balanced difference approximations to the gradients.

Now that we have the dynamic equations for the density matrix elements, we turn to determining the expression for the current density. Following the formalism of Aversa and Sipe \cite{aversa1995nonlinear} the current density can be expressed as 
\begin{equation}\mathbf{J}(t)=\frac{e}{m}\text{Tr}\left\{ \mathbf{p}\rho(t)\right\}.\end{equation} Using the fact that $\frac{\mathbf{p}}{m}=\frac{1}{i\hbar}\left[\mathbf{r},H\right],$ and decomposing the position operator into intraband and interband parts, $\mathbf{r}=\mathbf{r}_{i}+\mathbf{r}_{e},$ we can write this as \cite{aversa1995nonlinear}
\begin{equation}\begin{aligned}\mathbf{J}(t) & =\frac{e}{i\hbar}\text{Tr}\left\{ \left[\mathbf{r},H\right]\rho(t)\right\} \\
 & =\frac{e}{i\hbar}\sum_{n}\sum_{\mathbf{k}}\left\langle n,\mathbf{k}\right.|\left[\mathbf{r}_{i},H\right]\rho(t)\left|n,\mathbf{k}\right\rangle \\
 & +\frac{e}{i\hbar}\sum_{n}\sum_{\mathbf{k}}\left\langle n,\mathbf{k}\right.|\left[\mathbf{r}_{e},H\right]\rho(t)\left|n,\mathbf{k}\right\rangle, 
\end{aligned}\end{equation} where the trace is over the single electron states, and $\rho(t)$ is the reduced density matrix with matrix elements $\rho_{nm}(\mathbf{k})$. The decompostion of the position operator allows us to define the total current density as the sum of an intraband contribution, $\mathbf{J}_{i}$, and an interband contribution, $\mathbf{J}_{e}$. Using our expression for the Hamiltonian, as well as the matrix elements of the position operator (Eq. \ref{eq:1}), one may determine expressions for these contributions. The procedure is similar to that presented in recent work on MLG \cite{al2015nonperturbative}. After considerable work, the intraband current density near the $\mathbf{K}$-Dirac point can be shown to be given by

\begin{equation}\begin{aligned}\mathbf{J_{i}} & =\frac{\alpha e}{2\hbar}\sum_{\mathbf{k}}\frac{\mathbf{k}}{\tilde{\epsilon}}\left\{ \rho_{c_{1}c_{1}}(\mathbf{k})+\rho_{c_{2}c_{2}}(\mathbf{k})+\rho_{h_{1}h_{1}}(\mathbf{\mathbf{k}})+\rho_{h_{2}h_{2}}(\mathbf{\mathbf{k}})\right\} \\
 & -\frac{2e^{2}}{\hbar}\sum_{\mathbf{k}}\textrm{Re}\left\{ \rho_{c_{1}v_{2}}(\mathbf{k})\mathbf{\nabla_{\mathbf{k}}}\left(\mathbf{E}(t)\cdot\mathbf{\xi_{v_{2}c_{1}}}(\mathbf{\mathbf{k}})\right)\right.\\
 & +\rho_{c_{2}v_{1}}(\mathbf{k})\mathbf{\nabla_{\mathbf{k}}}\left(\mathbf{E}(t)\cdot\mathbf{\xi_{v_{1}c_{2}}}(\mathbf{\mathbf{k}})\right)\\
 & +\left[\rho_{c_{2}v_{2}}(\mathbf{\mathbf{k}})-\rho_{c_{1}v_{1}}(\mathbf{\mathbf{k}})\right]\mathbf{\nabla_{\mathbf{k}}}\left(\mathbf{E}(t)\cdot\mathbf{\xi_{v_{2}c_{2}}}(\mathbf{k})\right)\\
 & +\left.\left[\rho_{c_{2}c_{1}}(\mathbf{\mathbf{k}})+\rho_{v_{2}v_{1}}(\mathbf{k})\right]\mathbf{\nabla_{\mathbf{k}}}\left(\mathbf{E}(t)\cdot\mathbf{\xi_{c_{1}c_{2}}}(\mathbf{k})\right)\right\},
\end{aligned}\label{eq:3}\end{equation} while the interband current density is given by

\begin{equation}\begin{aligned}\mathbf{J}_{e} & =2e\sum_{\mathbf{k}}\text{Re}\left\{ \left(\frac{\tilde{\epsilon}-t_{\perp}}{2\tilde{\epsilon}}\right)\frac{\left(-k_{y}\hat{\mathbf{x}}+k_{x}\hat{\mathbf{y}}\right)}{k^{2}}\dot{\rho}_{c_{1}v_{2}}(\mathbf{k})\right.\\
 & +\left(\frac{\tilde{\epsilon}+t_{\perp}}{2\tilde{\epsilon}}\right)\frac{\left(-k_{y}\hat{\mathbf{x}}+k_{x}\hat{\mathbf{y}}\right)}{k^{2}}\dot{\rho}_{c_{2}v_{1}}(\mathbf{k})\\
 & +\frac{i\sqrt{\alpha}t_{\perp}}{2\tilde{\epsilon}^{2}}\frac{k_{x}\hat{\mathbf{x}}+k_{y}\hat{\mathbf{y}}}{k}\left(\dot{\rho}_{c_{1}v_{1}}(\mathbf{k})-\dot{\rho}_{c_{2}v_{2}}(\mathbf{\mathbf{k}})\right)\\
 & -\left.\frac{\sqrt{\alpha}}{2\tilde{\epsilon}}\frac{\left(-k_{y}\hat{\mathbf{x}}+k_{x}\hat{\mathbf{y}}\right)}{k}\left(\dot{\rho}_{c_{2}c_{1}}(\mathbf{k})+\dot{\rho}_{v_{2}v_{1}}(\mathbf{\mathbf{k}})\right)\right\}. 
\end{aligned}\label{eq:4}\end{equation} The sums over $\mathbf{k}$ in the current density expressions are restricted to a region near the $\mathbf{K}$-Dirac point, and we include a factor of 2 to account for spin degeneracy.
We also need to take into account the current density near the $\mathbf{K}^{\prime}$ point. Similar to MLG, we find that due to the symmetry of the Brillouin zone, the current densities around both Dirac points are identical. To obtain the total current density of BLG, we then only need to multiply the results calculated at the $\mathbf{K}$-point by two in order to account for this degeneracy. 

In what follows, we consider a suspended BLG sample, and use the time-dependent current densities to calculate the transmitted and the reflected THz fields, using a procedure identical to that used for MLG \cite{al2015nonperturbative}. We show in the next section that for low amplitude fields, we obtain the expected linear conductivities, which relate the THz field at the graphene sheet to the induced current densities. Furthermore, we have verified convergence in the nonlinear regime by altering the grid density, the extent of the grid, the time-step tolerance, and the polarization of the incident field.

\subsection{Linear Response}
In order to calculate the linear response of BLG to an incident field, we need to calculate expressions for the density matrix elements to first order in the THz field. Once we have these, we may then use Eq. (\ref{eq:3}) to express the first order \textit{intraband} current density as
\begin{equation}\begin{aligned}\mathbf{J_{i}}^{(1)} & =\end{aligned}
C(\omega_{p})\int_{0}^{\infty}d\tilde{\epsilon}^{\prime}\left(1+\frac{t_{\perp}^{2}}{\left(\tilde{\epsilon}^{\prime}+t_{\perp}\right)^{2}}\right)\sum_{n}f_{FD}(E_{n}),\label{eq:intra}\end{equation}
where $\tilde{\epsilon}^{\prime}$ is a shifted energy, given by $\tilde{\epsilon}^{\prime}=\tilde{\epsilon}-t_{\perp}$. $C(\omega_{p})$ is a frequency-dependent coefficient that includes the electric field: $$C(\omega_{p})=\frac{i|e|^{2}\mathbf{E}\left(\omega_{p}\right)e^{-i\omega_{p}t}}{2\hbar^{2}\pi\left(\omega_{p}+i/\tau_{c}\right)},$$ where we take our field to be harmonic, $$\mathbf{E}\left(t\right)=\mathbf{E}\left(\omega_{p}\right)e^{-i\omega_{p}t}.$$  In Eq. (\ref{eq:intra}), the sum is over the Fermi-Dirac distributions of the electrons and holes in the four bands of BLG. This is given expilicity by 
\begin{equation}\begin{aligned}\sum_{n}f_{FD}(E_{n}) & =\frac{1}{1+e^{\beta(\frac{\tilde{\epsilon}^{\prime}}{2}-\mu)}}+\frac{1}{1+e^{\beta(\frac{\tilde{\epsilon}^{\prime}}{2}+\mu)}}\\
 & +\frac{1}{1+e^{\beta(\frac{\tilde{\epsilon}^{\prime}}{2}+t_{\perp}-\mu)}}+\frac{1}{1+e^{\beta(\frac{\tilde{\epsilon}^{\prime}}{2}+t_{\perp}+\mu)}},
\end{aligned}\end{equation} where $\beta=k_{B}T$ and $\mu$ is the chemical potential of the system. 

As a check, we may evaluate the first order intraband current density in the limit of zero interlayer coupling. In this limit, $t_{\perp}\rightarrow0$, only the integral of the sum over the distributions remains. Thus our intraband current density reduces to
\begin{equation}\begin{aligned}\begin{aligned}\mathbf{J_{i}}^{(1)} & =C(\omega_{p})4k_{B}T\left\{ \text{ln}\biggl(1+e^{\beta\mu}\biggr)+\text{ln}\biggl(1+e^{-\beta\mu}\biggr)\right\} \\
 & =C(\omega_{p})8k_{B}T\text{ln}\left[2\text{Cosh}(\beta\mu/2)\right].
\end{aligned}\end{aligned}\end{equation} From here it is easy to read off the first order intraband conductivity as being
\begin{equation}\begin{aligned}\lim_{t_{\perp} \to 0}\sigma_{i}^{(1)}(\omega) & =\frac{4i|e|^{2}k_{B}T}{\hbar^{2}\pi\left(\omega_{p}+i/\tau_{c}\right)}\text{ln}\Bigl[2\text{Cosh}(\beta\mu/2)\Bigr].\end{aligned}\end{equation} 
This is the previously found result for the monolayer reponse \cite{mikhailovnonlin2009}, multiplied by two to account for the two layers of BLG. 

For BLG ($t_{\perp} \neq 0$), the integral in Eq. (\ref{eq:intra}) must be evaluated numerically in general, however in the limit that $T\rightarrow0$, it may be evaluated analytically. In this limit the Fermi-Dirac distributions will behave as Heaviside step functions; having the effect that only energies lower than the Fermi level, $E_{F}$, contribute to the integral. Assuming that our Fermi level is non-zero and located in the conduction bands, we only get contributions from the $\rho_{c_{1}c_{1}}\left(\mathbf{k}\right)$ and $\rho_{c_{2}c_{2}}\left(\mathbf{k}\right)$ distributions. In this case, we obtain two distinct contributions to the intraband current density. Thus, we are able to determine the intraband conductivities of both conduction bands in the limit of zero temperature. For the low-energy band $c_{1}$, this is given by \begin{equation} \lim_{T \to 0}\sigma_{i}^{(1)}(\omega)_{c_{1}}=\frac{i|e|^{2}E_{F}}{\hbar^{2}\pi\left(\omega_{p}+i/\tau_{c}\right)}\frac{(E_{F}+t_{\perp})}{(E_{F}+\frac{t_{\perp}}{2})},\end{equation} while for the high-energy band $c_{2}$, we have \begin{equation}\lim_{T \to 0}\sigma_{i}^{(1)}(\omega)_{c_{2}}=\frac{i|e|^{2}E_{F}}{\hbar^{2}\pi\left(\omega_{p}+i/\tau_{c}\right)}\frac{(E_{F}-t_{\perp})}{(E_{F}-\frac{t_{\perp}}{2})}\theta(E_{F}-t_{\perp}).\end{equation} Here the step function assures that the Fermi level must be greater than $t_{\perp}$ for $\rho_{c_{2}c_{2}}\left(\mathbf{k}\right)$ to provide a contribution.
 
Similarly, we may now use Eq. (\ref{eq:4}), and the connection elements from Eq. (\ref{eq:zeta}), to express the first order \textit{interband} current density as 
\begin{equation}\begin{aligned}\mathbf{J}_{e}^{(1)} & =D(\omega_{p})\int_{0}^{\infty}d\tilde{\epsilon}^{\prime}\left\{ \left[\frac{(\tilde{\epsilon}^{\prime}+2t_{\perp})}{\tilde{\epsilon}^{\prime}\left(\tilde{\epsilon}^{\prime}+t_{\perp}\right)\left(\frac{\tilde{\epsilon}^{\prime}}{\hbar}-\omega_{p}-i/\tau\right)}\right.\right.\\
 & +\frac{t_{\perp}^{2}}{\left(\tilde{\epsilon}^{\prime}+t_{\perp}\right)^{3}\left(\frac{\tilde{\epsilon}^{\prime}+t_{\perp}}{\hbar}-\omega_{p}-i/\tau\right)}\\
 & -\left.\frac{1}{\left(\tilde{\epsilon}^{\prime}+t_{\perp}\right)\left(\frac{t_{\perp}}{\hbar}-\omega_{p}-i/\tau\right)}\right]\times R(\frac{\tilde{\epsilon}^{\prime}}{2})\\
 & +\left[\frac{\tilde{\epsilon}^{\prime}}{\left(\tilde{\epsilon}^{\prime}+t_{\perp}\right)\left(\tilde{\epsilon}^{\prime}+2t_{\perp}\right)\left(\frac{\tilde{\epsilon}+2t_{\perp}}{\hbar}-\omega_{p}-i/\tau\right)}\right.\\
 & +\frac{t_{\perp}^{2}}{\left(\tilde{\epsilon}^{\prime}+t_{\perp}\right)^{3}\left(\frac{\tilde{\epsilon}^{\prime}+t_{\perp}}{\hbar}-\omega_{p}-i/\tau\right)}\\
 & +\left.\left.\frac{1}{\left(\tilde{\epsilon}^{\prime}+t_{\perp}\right)\left(\frac{t_{\perp}}{\hbar}-\omega_{p}-i/\tau\right)}\right]\times R(\frac{\tilde{\epsilon}^{\prime}}{2}+t_{\perp})\right\} +c.c,
\end{aligned}\label{eq:inter}\end{equation} where $D(\omega_{p})$ is a frequency dependent coefficient given by $$D(\omega_{p})=\frac{-i\omega_{p}|e|^{2}}{4\pi\hbar}\mathbf{E}\left(\omega_{p}\right)e^{-i\omega_{p}t},$$ and the function $R(x)$ is dependent on the temperature, Fermi level and energy of the system, which is given explcitly by $$R(x)=\frac{\sinh\beta(x)}{\cosh\beta\mu+\cosh\beta(x)}.$$ 

We may evaluate the above integral analytically in the limit of zero temperature and scattering, where $T \to 0$ and $\tau \to \infty$, respectively. In this limit, the complex factors in each term reduce to a Dirac delta function - when we neglect the contribution of the principal part of the integral. Taking only the positive frequency portion of the above expression, we obtain for the first order interband conductivity
\begin{equation}\begin{aligned}\lim_{\mathclap{\substack{T \to 0 \\ \tau \to \infty}}}\sigma_{e}^{(1)}(\omega_{p}) & =\frac{|e|^{2}}{4\hbar}\left\{ \frac{(\hbar\omega_{p}+2t_{\perp})}{\left(\hbar\omega_{p}+t_{\perp}\right)}+\frac{2t_{\perp}^{2}}{\hbar^{2}\omega_{p}^{2}}\theta\left(\hbar\omega_{p}-t_{\perp}\right)\right.\\
 & +\left.\frac{\left(\hbar\omega_{p}-2t_{\perp}\right)}{\left(\hbar\omega_{p}-t_{\perp}\right)}\theta\left(\hbar\omega_{p}-2t_{\perp}\right)\right\}. 
\end{aligned}\end{equation} This expression agrees with those found in the literature \cite{abergel2007optical, nicol2008optical}. We see that BLG exhibits a minimum conductivity that is associated with interband transitions. For very large incident frequencies, this minimum conductivity is given by $\frac{|e|^{2}}{2\hbar}$, which is twice that found in monolayer graphene, as expected. 

We may also use Eqs. (\ref{eq:intra}) and (\ref{eq:inter}) to determine the first order current densities at non-zero temperatures and scattering times, which is a closer approximation of experimental conditions. We present the results of this analysis in the next section.

\section{Simulation Results}\label{sec:3}
We have shown that our analytic expressions for the linear THz response agrees with the literature in certain limits. Going beyond the linear regime to determine the response at higher order in the THz field is our main interest. We want to be able to examine the dependency of the THz response on the temperature and incident frequency of BLG. To achieve this aim we employ a simulation, which offers the ability to control the desired parameters, and calculate the interband and intraband current densities at the desired field amplitudes. The field transmitted from the BLG is calculated as a function of the current densities and the incident field. The transmitted field is then spectrally analyzed to determine its frequency components; high harmonic generation in the spectral composition indicating nonlinear behaviour.

We begin with a check on the first order calculations by comparing the linear conductivity derived from the first order equations of the previous section to that calculated by the full simulation. We then proceed to examine the higher order response of BLG, determining at which incident field amplitudes we might expect to see the largest generation of the third harmonic. 

\subsection{Linear Results}
\begin{figure}[h]
\centering
\includegraphics[width=0.5\textwidth]{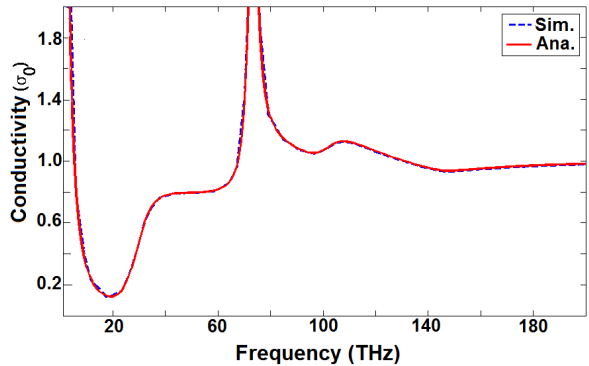}
\caption{Comparison of the full linear conductivity, calculated by computer simulation and numerical integration of the closed form expression. Fermi level is $60\, meV\, (0.2t_{\perp})$, scattering time is $80\, fs$, and temperature is 50 K. Conductivity is measured in units of the universal conductivity of BLG, $\sigma_{0}=e^{2}/2\hbar$\label{fig:4.2}.}
\end{figure}

\begin{figure}[h]
\centering
\includegraphics[width=0.5\textwidth]{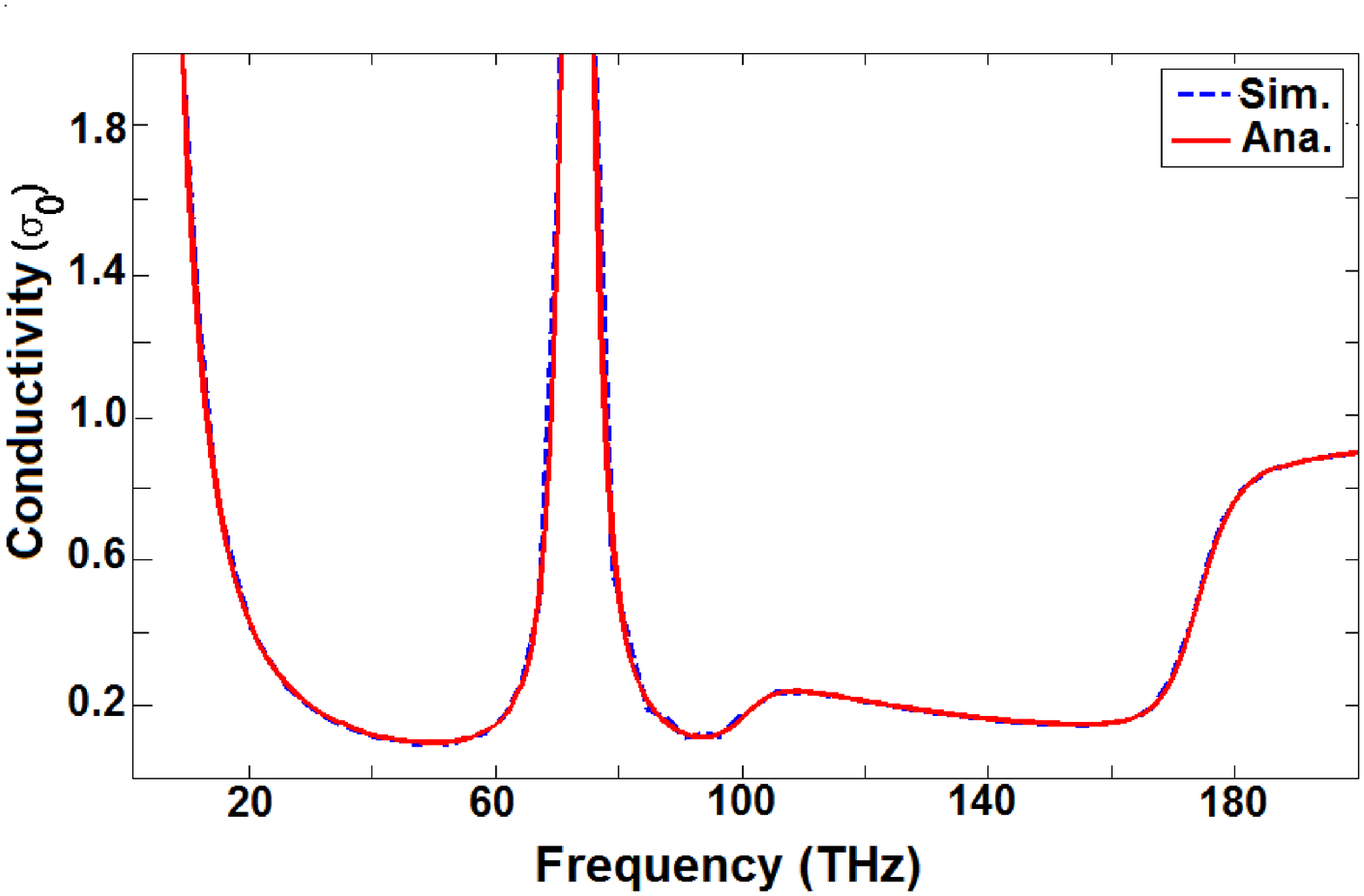}
\caption{Comparison of the full linear conductivity, calculated by computer simulation and numerical integration of the closed form expression. Fermi level is $360\, meV\, (1.2t_{\perp})$, scattering time is $80\, fs$, and temperature is 50 K. Conductivity is measured in units of the universal conductivity of BLG, $\sigma_{0}=e^{2}/2\hbar$\label{fig:4.2-1}.}
\end{figure}

As an initial check, we compare on a single plot the conductivity arising from the full current density (intraband and interband), as calculated by both the simulation and the closed form expressions (Eqs. (\ref{eq:intra}) and (\ref{eq:inter})). We plot these results in Figs. \ref{fig:4.2} and \ref{fig:4.2-1} for Fermi levels of 60 and 360 meV respectively, which correspond to energies of $0.2t_{\perp}$ and $1.2t_{\perp}$. These values also allow us to make comparisons to similar work found in the literature \cite{nicol2008optical}. In both instances, the scattering time is $80\, fs$, and the temperature is 50 K. As can be seen the agreement between the simulated and first order results is excellent.

In both plots we see features at incident freqeuncies of $f=0$ and $f\simeq75$ THz. The zero frequency peak is a result of the intraband Drude response. This feature vanishes for zero Fermi level and temperature, as the intraband transitions become negligible in this limit. The second peak at $f\simeq 75$ THz is due to the difference in energy between the two conduction bands of BLG ($\simeq300\, meV$). Specifically, it arises from the perfect nesting between the dispersions of the higher and lower conduction bands, $c_{2}\,\&\, c_{1}$. These two bands sit directly on top of one another and differ in energy only by the constant displacement $t_{\perp}$. Thus, the nesting of the bands leads to a large spectral weight at frequencies close to $t_{\perp}$, which is approximately equivalent to 75 THz. The additional features present in the plots reflect other possible interband transitions, such as $v_{1}\rightarrow c_{1}$ or $v_{1}\rightarrow c_{2}$. These transitions require incident radiation of energy $t_{\perp}$and $2t_{\perp}$, respectively, and their spectral weights are largely dependent on the Fermi level.

\subsection{Nonlinear Results}
We now present our simulation results for the nonlinear THz response of BLG. In each of our simulations for BLG we took the sample to be undoped $(\mu=0)$. The response was then calculated at field amplitudes ranging from 0.5 kV/cm to 2.0 kV/cm at two different temperatures: 10 K and 100 K.  This is followed by the calculation of the response at several incident frequencies at a temperature of 10 K. For the system and fields modelled in this section, only the $v_{2}$ and $c_{1}$ bands contribute to the THz response.

Our input THz field is a sinusoidal Gaussian pulse with central frequency of 1 THz and full width at half maximum (FWHM) of 1 ps. Mathematically, this may be expressed as
\begin{equation}
\mathbf{E}_{i}(t)=\mathbf{E}_{0}e^{-4\textrm{log}(2)\left(\frac{t-t_{0}}{T_{FWHM}}\right)^{2}}\textrm{sin}\left[2\pi f_{0}(t-t_{0})\right],
 \end{equation}
where $t_{0}$ and $T_{FWHM}$ are the temporal shift and full width at half maximum of the Gaussian pulse, respectively. The central frequency of the pulse is given by $f_{0}$. We take our scattering and relaxation times to be fixed at 50 fs, which is an average of theoretical and measured values for such (sample-dependent) constants. This is a conservative estimate for undoped BLG at low temperature; an increase in this time will lead to stronger harmonic generation \cite{al2014high}.  In the simulation, we keep the Fermi level at the Dirac point and as carriers are injected we raise the temperature of the distribution to which the carriers relax to account for the increase in the carrier density.

\subsubsection*{T=10 K Results}
At a temperature of 10 K, the intrinsic thermal carrier density in BLG is approximately $1.4\times10^{10}/cm^{2}$ and the electrons have an average energy of 1.03 meV. In comparison, at this temperature MLG has an intrinsic carrier density of approximately $2.0\times10^{8}/cm^{2}$, and an average electron energy of 1.89 meV. Thus, we find a much larger thermal carrier density in BLG than we do in MLG. However, the average energy of the carriers is slightly larger in MLG. In both cases, the average electron energy is much less than the central photon energy of 4.14 meV associated with a 1 THz pulse, thus to first order in the field, interband transitions are largely unaffected by the thermal energy.

All of the carriers (injected and thermal) are driven within the conduction and valence bands by the applied electric field. The subsequent motion leads to a blocking and unblocking of the electronic states that are available for extra carrier injection. This results in an interplay between intraband and interband dynamics that is paramount in producing the nonlinear response \cite{al2014high}.

\begin{figure}[h]
\centering
\includegraphics[width=0.5\textwidth]{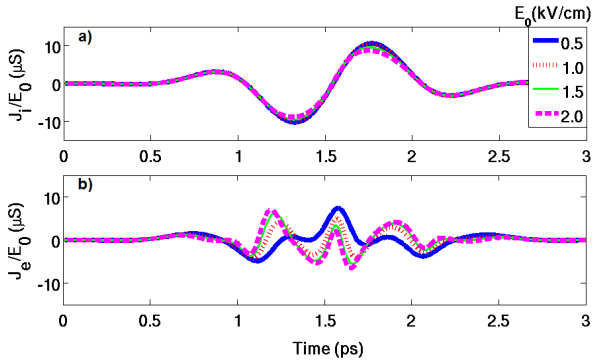}
\caption{Response of BLG to a incident field of 1 THz at T=10K and for four different field amplitudes. a) Intraband current density normalized to incident field amplitude. b) Interband current density normalized to incident field amplitude.}\label{fig:4.3}
\end{figure}

We now present results for a number of different field amplitudes. 
In Fig. \ref{fig:4.3}, we plot the intraband and interband current densities for four different incident field amplitudes (0.5, 1, 1.5 and 2 kV/cm). All current densities are normalized to the peak value, $E_{0},$ of the incident field such that, if the response were linear, these relative currents would be unchanged by an increase in incident field. This allows for a comparison of the currents at each field amplitude, and for the clear identification of any nonlinear behaviour. In what follows, we refer to these as \textit{relative current densities}. The calculated relative intraband and interband current densities at these specific field amplitudes are shown in Figs. \ref{fig:4.3}(a) and \ref{fig:4.3}(b), respectively. 

Looking first at the intraband current density, we find that as the field amplitude is increased, the relative intraband current decreases. This is due to the `clipping` phenomena that is a property of linear dispersions, such as that found in MLG. Although BLG has a parabolic dispersion near the Dirac point, as we move away from this point, the dispersion becomes linear (see Fig. \ref{fig:1}). Thus, the high field amplitudes drive carriers far enough away from the Dirac point such that they occupy the linear part of the band structure. In this region, the carrier velocities become constant and the field is not able to drive them to higher velocities; the intraband current is clipped as a result. 

We note, however, that the current clipping nonlinearity in the intraband current density is much smaller than the nonlinearity seen in the interband current. We also note that due to the large intrinsic carrier density of BLG due to the large number of states, carrier injection has only a minor effect on the intraband current density at these field amplitudes. This is in contrast to MLG, where one finds a substantial increase in the relative intraband current density arising from the injection of carriers \cite{al2014high}. 

\begin{figure}[h]
\centering
\includegraphics[width=0.5\textwidth]{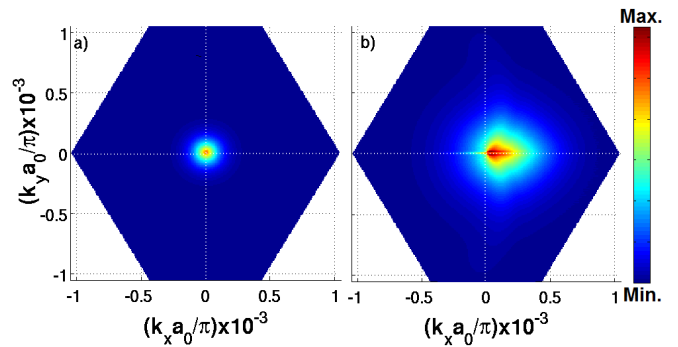}
\caption{Electron density distribution in k-space (a) for the initial thermal distribution and (b) at time t=1.75 ps for a 0.5 kV/cm incident field. White lines indicate the position of the Dirac point.}\label{fig:density}
\end{figure}

Examining next the interband current density, we observe large distortions for all of the field amplitudes. Based on the strength of these distortions, we expect the interband current to be a large source of nonlinearity at these field amplitudes. The motion of the charge carriers in the conduction band gives rise to an interplay between intraband and interband dynamics. We can visualize this interplay between interband and intraband motion via a plot of the electron density in the conduction band, as shown in Fig. \ref{fig:density}. 
As the pulse hits the BLG sample, the population of carriers in the conduction band is driven away from the Dirac point by the incident THz field; this driving of the carriers within each band is the source of the intraband current. Fig. \ref{fig:density}(a) shows the initial electron density before the pulse hits. In Fig. \ref{fig:density}(b) we see the density at t = 1.75 ps, where the magnitude of the amplitude of the incident field is a maximum. We can see the distribution has been driven to the right of the Dirac point. This intraband motion also opens up electron states near the Dirac point that were previously occupied before the pulse arrived. As a result of the reduction in Pauli blocking, carriers from the valence band $v_{2}$ may be injected into the conduction band $c_{1}$ near the Dirac point. This is evidenced in the dark, high-density regions that appear above and below the $k_{y}=0$ line to the right of the Dirac point in Fig. \ref{fig:density}(b). This carrier injection is the source of the interband current.

Thus, we see that there is an interesting relationship between the intraband and interband dynamics; the motion of carriers within each band has an effect on both interband and intraband current densities. Importantly, this relationship manifests itself in the nonlinear response of BLG.

\begin{figure}[h]
\centering
\includegraphics[width=0.5\textwidth]{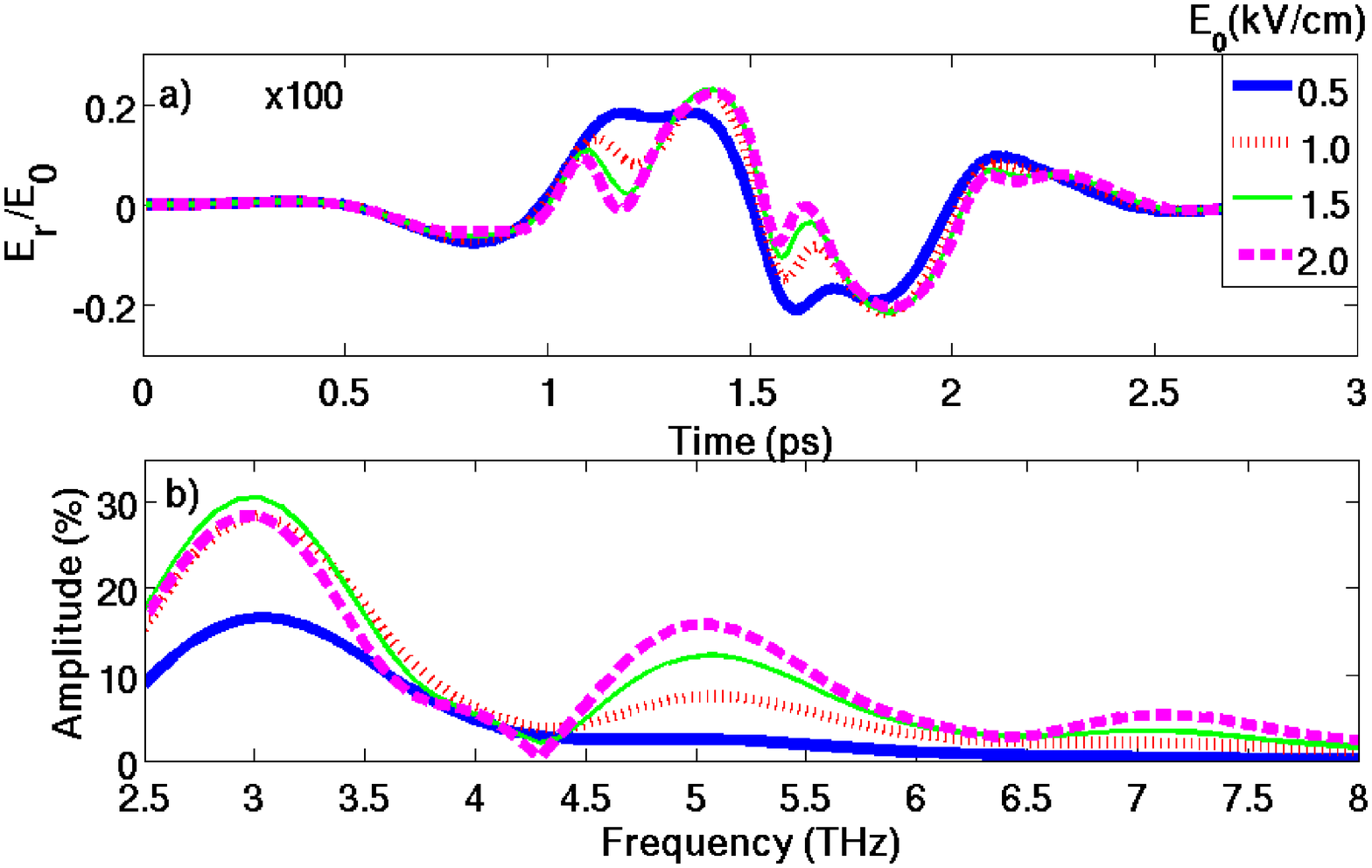}
\caption{Response of BLG to a incident field of 1 THz at T=10K for four different field amplitudes. a) The reflected field in the temporal domain, normalized to the amplitude of the incident field. Value of $E_r/E_0$ is multiplied by 100 for clarity. b) The amplitude spectra of the reflected signal normalized to the peak at the fundamental frequency of 1 THz.}\label{fig:4.3-1}
\end{figure}

We now plot the time-dependent reflected fields normalized to the peak value of the different incident field amplitudes in Fig. \ref{fig:4.3-1}(a), and the spectral responses normalized to the peak amplitude at the fundamental frequency (1 THz) in Fig. \ref{fig:4.3-1}(b). 

In the upper plot we can see strong distortion of the reflected fields for all of the selected field amplitudes. Note that because there is no substrate the induced current is the entire source of the reflected field, which only has a peak amplitude of approximately 2 V/cm for the incident field of 1 kV/cm. To see harmonic content in the reflected field, Fig. \ref{fig:4.3-1}(b) we plot the Fourier transform of the reflected fields normalized to the peak reflected field at the fundamental frequency. We find that at the lowest field amplitude of 0.5 kV/cm, there is a clear third harmonic signal. Furthermore, we see a maximum in the third harmonic of the relfected field at an incident field amplitude of about 1.5 kV/cm. At this amplitude the third harmonic reaches a maximum of approximately 30\% of the reflected spectral peak at the fundamental. As we increase the incident field amplitude further, we begin to see a decrease in the third harmonic. 

Moreover, at these high field amplitudes we also see the presence of the 5th harmonic, which is largely absent at lower field amplitudes. Its amplitude takes a maximum value of approximately 15\% the spectral peak at the fundamental for an incident field of 2 kV/cm. The appearance of the fifth harmonic indicates that at the higher field amplitudes fifth order processes are significant.  Just as was found with MLG \cite{al2014high}, the fifth order response results in a decrease in the third harmonic.

\subsubsection*{T=100 K Results}
We next present results for simulations at a temperature of 100 K. At this temperature, the intrinsic thermal carrier density is $3.4\times10^{10}/cm^{2}$ and the average thermal electron energy is 10.8 meV. This is compared to the average photon energy of 4.14 meV associated with a 1 THz pulse. Thus, the Pauli blocking of the interband transitions is increased due to the increased thermal populations of carriers; reducing the overall interband current density. Additionally, the increase in thermal carriers gives rise to an increase in the intraband current density. Thus, at this temperature the intraband current is dominant, resulting in a reduction in the interplay between intraband and interband dynamics, and ultimately a reduction in high harmonic generation. 
%\begin{figure}[h]
%\centering
%\includegraphics[width=0.5\textwidth]{Figures/Chapter4/paper1T=100.eps}
%\caption{Response of BLG to a incident field of 1 THz for four differnt field amplitudes and T=100K. Top) Interband current density normalized to incident field amplitude. Bottom) Intraband current density normalized to incident field amplitude.}
%\end{figure}

\begin{figure}[h]
\centering
\includegraphics[width=0.5\textwidth]{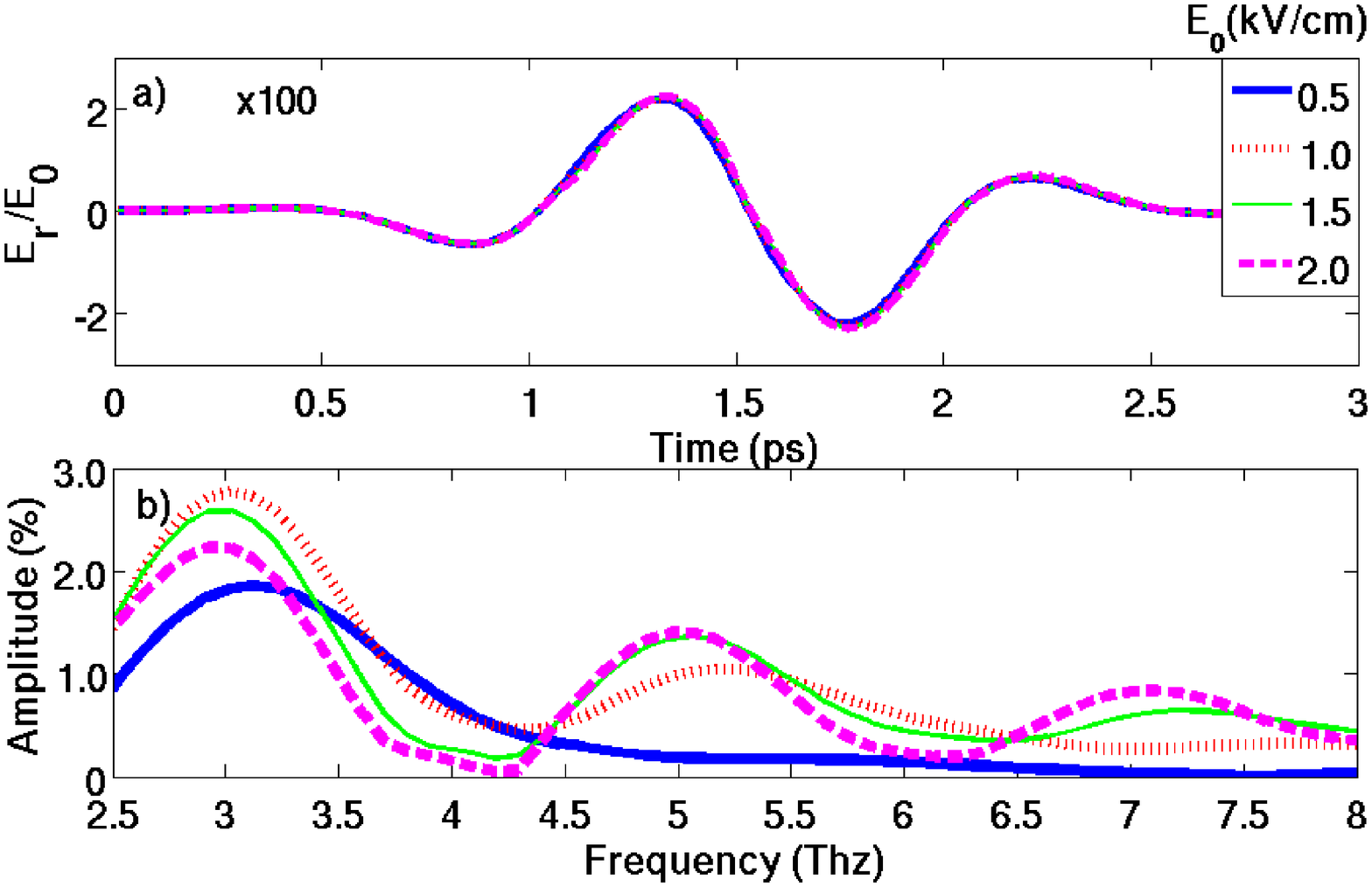}
\caption{Response of BLG to a incident field of 1 THz at T=100K for four different field amplitudes. a) The reflected field in the temporal domain, normalized to the amplitude of the incident field. Value of $E_r/E_0$ is multiplied by 100 for clarity. b) The amplitude spectra of the reflected signal normalized to the peak at the fundamental frequency of 1 THz.}\label{fig:4.4}
\end{figure}

We can see this by looking again at both the reflected field and the spectral response for each of the incident field amplitudes. The normalized time-dependent reflected fields for the different field amplitudes are shown in Fig. \ref{fig:4.4}(a), and the spectral responses normalized to the peak amplitude at the fundamental frequency (1 THz) are presented in Fig. \ref{fig:4.4}(b). 

In the upper plot we see far less distortion in the reflected fields than we do at T=10 K, for all of the selected field amplitudes. The reduction in distortion is again due to the increase in the thermal carrier density and subsequent increase in Pauli blocking. In turn this results in diminished harmonic generation,  which is shown in Fig. \ref{fig:4.4}(b). We find a maximum in the third harmonic of the relfected field at an incident field amplitude that is now approximately 1 kV/cm. At this amplitude it reaches a maximum of only approximately 2.8\% of the reflected peak at the fundamental, which is still greater than that found in MLG at the same temperature \cite{al2015nonperturbative}. At higher field amplitudes, we also see the presence of the 5th harmonic.  Its amplitude takes a maximum value of approximately 1.5\% of the peak at the fundamental for an incident field of 2 kV/cm. 

Thus, we find that as the temperature of the system is increased, the nonlinear response - specifically, high harmonic generation - is greatly diminished as a result of the reduction in the interplay between interband and intraband dynamics. Therefore, if one is to observe high harmonic generation (HHG) experimentally in BLG, it is apparent that low temperatures are necessary. This allows for the maximization of the dynamic interplay that is paramount to the nonlinear repsonse. 

\subsubsection*{Effect of Central Frequency}
\begin{figure}
\centering
\includegraphics[width=0.5\textwidth]{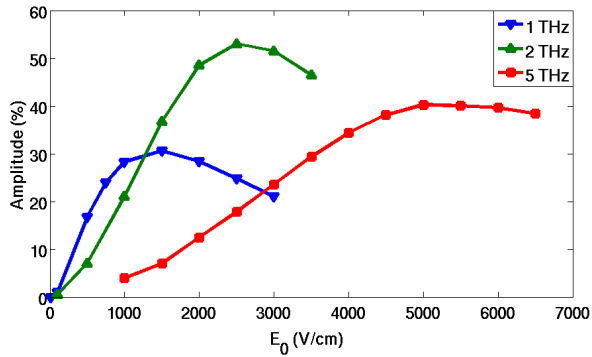}
\caption{Third harmonic amplitude spectra normalized to the peak of the fundamental central frequency, for three different incident frequencies: 1,\,2 and 5 THz.}\label{fig:4.5}
\end{figure}

Finally, we study the effect that the central frequency of the incident pulse has on the generation of the third harmonic. The interplay between intraband and interband dynamics has a strong dependence on the central frequency, because when the central frequency is increased, the carriers are injected farther from the Dirac point. In Fig.  \ref{fig:4.5} , we plot the normalized third harmonic amplitude as a function of the incident THz field amplitude for central frequencies of 1.0, 2.0 and 5.0 THz. In each case, we adjust the duration of the pulse such that the product of the central frequency and the pulse duration is constant, so that the pulse remains single-cycle for each of the central frequencies considered.

As the central frequency of the incident pulse is increased, we observe a maximum third harmonic amplitude that is larger in comparison to the simulations performed at 1 THz. For a central frequency of 2 THz, the normalized third harmonic amplitude peaks at approximately 53\% of the fundamental for a field amplitude of 2.5 kV/cm. This is compared to a value of 30\% for the central frequency of 1 THz at a field amplitude of 1.5 kV/cm.

As the pulse frequency is increased further to 5 THz, the maximum normalized third harmonic is found to decrease to 40\% of the fundamental. Furthermore, this maximum value is not obtained until the input field amplitude reaches 5 kV/cm. 

The observed increase in the nonlinear interplay with increasing central frequency is a result of a number of factors. First, due to the approximately inverse relationship between the linear conductivity and the frequency, the intraband current density decreases as we increase the central frequency of the pulse. This in turn increases the ratio of the third harmonic to the fundamental, due to the reduction in the part of the reflected fundamental that arises due to the intraband current. 

Additionally, we find that as the incident frequency is increased there is an increase in the interband current density. The increase arises because the interband transitions that are resonant with the 2 THz and 5 Thz photons are mostly unoccupied when the pulse arrives, whereas at lower frequencies they may be occupied by either thermal or injected carriers. This larger interband current density results in an increased  dynamic interplay, and thereby yields a larger third harmonic field.

\vspace{5mm} 
\section{Summary}\label{sec:4}
We have presented a detailed study of the nonlinear response of unbiased bilayer graphene at THz frequencies. A theoretical model has been developed, which is based on the dynamic equations of density matrix elements, employing the length gauge. The model enabled the calculation of eigenvalues and eigenvectors of BLG, as well as interband and intraband connection elements. Expressions for interband and intraband current densities were also derived.

Through the use of simulation, we determined solutions of the density matrix dynamic equations. These solutions were then applied to the study of high harmonic generation in BLG. Investigating the effect of system parameters - such as the central THz frequency and the ambient temperature - on third harmonic generation was the main focus. 

Our results show that for a temperature of 10 K, scattering time of 50\,fs and incident field of 1 THz, the third harmonic can be as large as 30\% of the fundamental for an incident THz field of 1.5 \textrm{kV/cm}. As we increase the temperature to 100 K, we find that the maximum third harmonic generation is reduced by an order of magnitude, due to the reduction in interband transitions. 

Finally, we found that as the central THz frequency is increased from 1 to 5 THz, we see an increase in the third harmonic amplitude; reaching a maximum of 53\% of the fundamental for a 2 THz incident field at 2.5 kV/cm. These results may be of use to experimentalists aiming to probe the nonlinear response of bilayer graphene. 

To experimentally observe the high harmonics we predict for BLG, one must consider the dynamic range of the THz spectrometers - defined as the ratio of the frequency dependent signal strength to the detected noise floor \cite{jepsen2005dynamic}. For a 1 THz  incident field of 1 kV/cm, we find the peak amplitude of the reflected field to be approximately 2.2 V/cm (53 dB less than the incident field). Thus, a detection technique that allows for a dynamic range larger than 53 dB is required for the measurement of the reflected signal. Such a dynamic range is experimentally feasible; a dynamic range of 90 dB has been reported in recent work \cite{vieweg2014terahertz}. 

\bibliographystyle{apsrev4-1}

\bibliography{UCThesisBibliography}

\end{document}